%% file: mad.tex
\documentclass[12pt]{iopart}
\usepackage{iopams}
\usepackage{epsfig}

\def\ead#1{\vspace*{5pt}\address{E-mail: \mailto{#1}}}
\def\mailto#1{{\tt #1}}
\input{comdef}

\begin{document}

\title
%\title[Determination of the longitudinal structure function $F_{L}$ at HERA]
{Selected Results on the Transition from Short to Long Distance
  Physics at HERA}

\author{J\"org Gayler  
}

\address{DESY, Notkestrasse 85, 22603 Hamburg, Germany}
\ead{gayler@mail.desy.de}

\begin{abstract}
 Recent results from the HERA $ep$ collider are  discussed with
     emphasis on the transition from short to long distance
     phenomena in QCD. The results cover inclusive $ep$ scattering,
     inclusive diffractive scattering, vector meson production,
      and deeply virtual Compton scattering (DVCS).
\end{abstract}



%\maketitle

\section{Introduction}
In electron proton ($ep$) scattering at high energies,
a wide spectrum
of phenomena can be explored which are in principle covered by
Quantum Chromodynamics (QCD),
% - the theory of the interactions between quarks and gluons -,
but which can not be described yet by rigorous detailed calculations.
 Whereas scattering processes involving hard interactions,
 such as production of jets, 
 can be accessed by perturbative QCD (pQCD),
 large distance phenomena
 corresponding to soft interactions and involving only small scales,
 are mainly explored by experiment. Even the results on the
 proton structure function $F_{2}$, which are often considered as
 a triumph of QCD,
   provide both, a measurement
  of parton densities (pdf) at low scales $Q^2$ and indeed a successful pQCD
test
 by the analysis of their evolution towards large $Q^2$.
%  (see~\cite{Adloff:2001qk,Chekanov:2001qu}
%  for recent precision results on $F_2$).
Here $Q^{2} \equiv -q^2$ is the 4-momentum squared of the exchanged
virtual photon with 
 $q=p_{e}-p_{e}^{\prime}$, where $p_{e}$ and $p_{e}^{\prime}$ are
the
  4-momenta of the incident and scattered electron respectively.

I present\footnote{talk presented at MAD'01, Antananarivo, October 2001}
% I present
 results on $F_2$
 which are relevant in context
 of the discussions on saturation
  (see~\cite{Gribov:1981ac} and e.g.~\cite{Navelet:1994qp}),
  i. e. the damping
 of the rise of $F_{2}$ at high energies (small $x$) which is expected
 due to general arguments like unitarity,
 or more specifically in context of QCD due to increasing parton densities
 at high energies and
 gluon self interactions.
 Of particular interest are here
 the transition
 from deep inelastic scattering (DIS)
 to photo production
 ($Q^2 \approx 0$) and the high
 energy limit at fixed $Q^{2} \neq 0$.

  Diffractive scattering,
  where the proton or photon (or dissociations of them)
  appear in the final state,
  is particularly interesting for the study of the
  transitions from short to long distance phenomena.
At high momentum transfer or in case of jet production,
such $ep$ interactions
  may be hard enough
to justify pQCD calculations.
   On the other hand, the frequency of
   the reappearance of the initial state
particles,
   in particular the proton, tells
  about the confining forces (confer the general
discussion in~\cite{Bartels:2001ze}).
  Here I present results
  on inclusive diffractive $ep$ scattering
  and also on specific final states where the virtual
photon,
  emitted from the electron and interacting with the proton,
  is transformed into a particle with photon quantum numbers, like
  a real photon or a vector meson.
  Such quasi elastic reactions are in close relation to fully inclusive
  scattering, i.e. to the $F_{2}$ structure function, via the optical
theorem.

 Many other results on non-perturbative phenomena are available from
 HERA and
 could not be covered
 in this talk,  like search for instantons, search for
 odderons~\cite{stiewe}, data on the photon structure, fragmentation,
 pion exchange and more.

% This paper contains in section 2 results on $F_{2}$, followed by a
% discussion
% of results on diffractive inclusive
% scattering in section 3. After some discussion of exclusive
% vector meson production in section 4, results on exclusive production
%of
% photons, i. e.  deeply virtual Compton scattering (DVCS), are
%presented
% in section 5, followed by conclusions.

\section{Inclusive reaction ($ep \rightarrow eX$)}
\begin{figure}[ht]
 \begin{center}
 \begin{picture}(200,274)
%      274 vor cosmetics
   \put(-7,0.){\epsfig{file=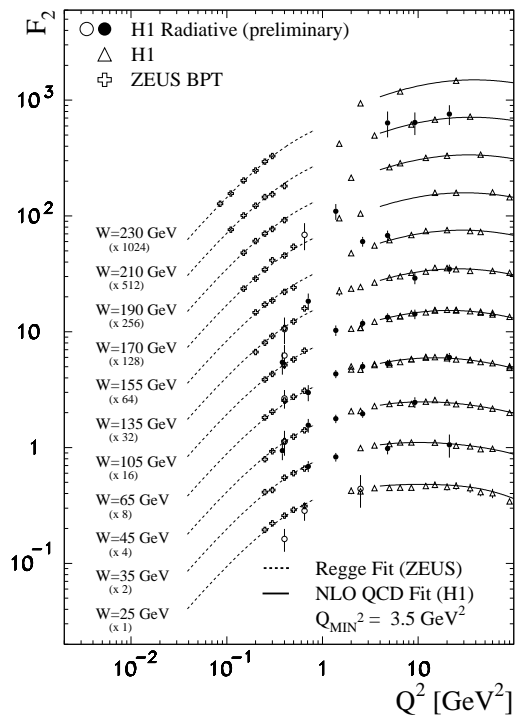,width=210pt}}
%    210 vor cosmetics
 \end{picture}
\caption{$F_{2}$ as function of $Q^{2}$ 
for different $W$.
\label{fig:f2lowq}}
 \end{center}
\end{figure}

The proton structure function $F_{2}$ follows in a wide
kinematic range the patterns expected from pQCD. In particular
the $Q^{2}$ dependence can be described by DGLAP evolution
%~\cite{dglap}
of pdfs which are parametrised at low scales
 consistently with
 experimental data (e.g.~\cite{Adloff:2001qk,Chekanov:2001qu,Martin:2001es}).
 However the measurements extend into the region of
 small $Q^{2}$ which correspond to transverse sizes of the
 interaction region well above 1 fm. Fig. \ref{fig:f2lowq} shows
 recent data~\cite{H1Zlowq,Breitweg:2000yn} on $F_{2}$ as function of $Q^{2}$
 for different $W$, the invariant mass of the hadronic final state,
 with $W^{2}=(p+q)^{2}$, where $p$ is the 4-momentum of the incident
proton.
  The region of low $Q^{2}$ is described
  by phenomenological models (here~\cite{Breitweg:2000yn}).
%  only, in particular the moderate rise of $F_{2}$ with energy
%  $W$ follows the expectation from Regge models~\cite{zeusregge}.
  In the limiting case of photo production ($Q^{2}=0$), the energy
  dependence is well described by Regge based models where the
 dependence on the total energy $W$ of the hadronic system
%  energy dependence
 is directly taken from soft hadronic
  reactions~\cite{Donnachie:1992ny}.

  That $F_{2} \rightarrow 0$  for $Q^{2} \rightarrow 0$, as suggested
  by Fig.~\ref {fig:f2lowq}, can be expected from the relation
\begin{equation} \label{eq:f2}
     F_{2} = 1/4\pi\alpha^{2} \; Q^{2}
     \sigma_{tot}^{\gamma^{*}p}
\end{equation}
 with the total virtual photo production cross section
 $\sigma_{tot}^{\gamma^{*}p}$,
 which is constant for $Q^{2}\rightarrow 0$.

\begin{figure}[ht]
 \begin{center}
\begin{picture}(200,236) 
%  vor cosmetic    236
   \put(-35,0.){\epsfig{file=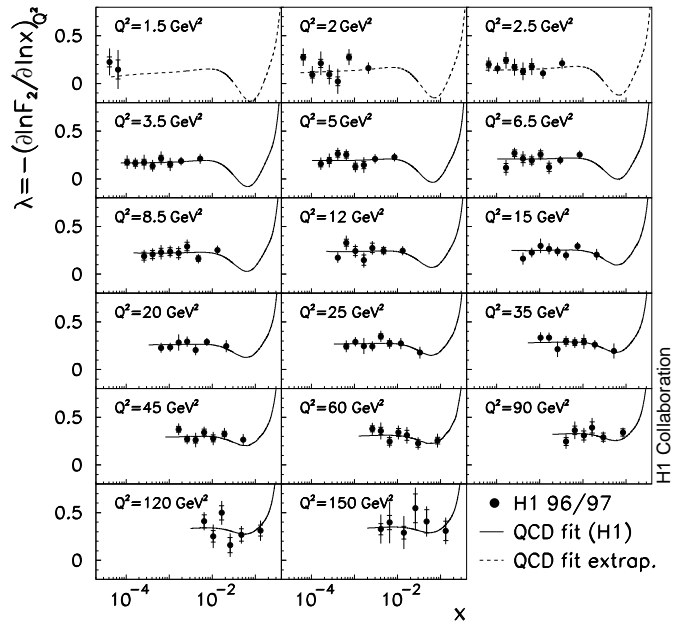,width=250pt}}
%   vor cosmetics   250
 \end{picture}
\caption{
 Derivative $\lambda=-(\partial \ln F_2/\partial \ln x)_{Q^2}$ 
%for different $Q^{2}$ with QCD analysis of ref.~\cite{Adloff:2001qk}.
 for different $Q^{2}$ with QCD analysis of~[5].
%     hand ref
\label{fig:rise}}
 \end{center}
\end{figure}

  In contrast to the smooth energy dependence of $\sigma_{tot}^{\gamma
  p}$, the energy dependence of $F_{2}$ is steep
  as demonstrated recently
% in Fig.~\ref{fig:xdep}
  with high precision~\cite{Adloff:2001qk,Chekanov:2001qu}
  in terms of the $x$ dependence of $F_{2}$.\footnote{
 The Bjorken variable  $x=Q^{2}/2pq$ is related to $W$ by
  $W^{2}=Q^{2}(1/x-1)$.}
  It is expected that this rise towards low $x$, which
  in pQCD analyses
  corresponds to an increase of the gluon density
 (see e.g.~\cite{Adloff:2001qk,Martin:2001es}),
  will slow down at
  highest energies (small $x$) due to gluon-gluon
  interactions~\cite{Gribov:1981ac}. H1 studied the $x$ dependence
  in detail and evaluated~\cite{Adloff:2001rw}
  the derivative $\lambda=-(\partial \ln F_2/\partial \ln x)_{Q^2}$ using
  recent
  $F_{2}$ data~\cite{Adloff:2001qk}.
  Fig.~\ref{fig:rise} shows that $\lambda$ is constant
  within
  experimental uncertainties for fixed $Q^{2}$ in the range $x < 0.01$,
  implying that the data are consistent with the power behaviour
  $F_{2} = \sim x^{-\lambda(Q^{2})}$.

  We conclude that no damping of the rise of $F_{2}$ is visible yet at
present energies and $Q^{2} \gtrsim 2$ GeV$^{2}$.
  In fact a very simple parametrisation
\begin{equation} \label{eq:f2p}
 F_2 = c \cdot x^{-\lambda(Q^2)}\;,\;\; {\rm with} \;\;
         \lambda(Q^2) = a \cdot \ln[Q^2/\Lambda^2]
\end{equation}
with $c \approx 0.18$
  describes the  low $x$ $F_{2}$ data very well.
% (see Fig.~\ref{fig:f2parms}).
%\begin{figure}[t]
%\begin{center}
%\epsfxsize=21pc % will enlarge or reduce the postscript figures
%\epsfbox{figs/c.eps} % postscript image file name
%\caption{ c
%  \label{fig:c}}
%\end {center}
%\end{figure}
%\begin{figure}
%\begin{center}
%\epsfxsize=21pc % will enlarge or reduce the postscript figures
%\epsfbox{figs/f2lam.eps} % postscript image file name
%\caption{
%    $\lambda$  \label{fig:f2lam}}
%\end {center}
%\end{figure}
  A fit~\cite{Adloff:2001rw} to the H1 data~\cite{Adloff:2001qk}
  at $x < 0.01$, $Q^{2} > 3.5$ GeV$^{2}$ yields
$a = 0.0481 \pm .0013 \pm .0037$,
 $\Lambda = 292 \pm 20 \pm 51$ MeV.

  At very low $Q^{2}$ a flattening of the $Q^{2}$ dependence of
  $\lambda$
  with $\lambda \approx 0.08$ at $Q^{2} \approx 0$
  is expected from relation (\ref{eq:f2})
 considering the energy dependence 
  of the total photo production cross section.
  The latter can be written as
  $\sigma_{tot}^{\gamma p} \sim s^{\alpha_{\pom}(0)-1}$
  where $\alpha_{\pom}(0)$,
  with $\alpha_{\pom}(0) -1 \approx 0.08$,
  is the intercept of the pomeron trajectory
  invented to describe the
  high energy dependence of soft hadronic
interactions~\cite{Donnachie:1992ny}.
  Here $s$ is the total centre of mass energy squared with
  $s = W^{2} \sim Q^{2}/x$.
  The approach of $\lambda$ to the soft pomeron intercept
  at $Q^{2} < 1$GeV$^2$ is indeed
  observed in an analysis
  by the ZEUS collaboration~\cite{zeuslambda}.
%  (see Fig.~\ref{zeuslam:fig}).

  Also the derivative $(\partial F_2/\partial \ln Q^2)_x$
  was previously discussed in context of the
  rise of $F_{2}$. If presented as function of
  $Q^{2} > 3$ GeV$^{2}$, a rise towards small $x$
  is observed~\cite{Adloff:2001qk}
  consistent with pQCD DGLAP analyses.
  In contrast, if presented for fixed $W$, a decrease is
observed~\cite{zeuslambda}
  versus small $x$ which is however kinematically related to a
  decrease of $Q^{2}$.

  In summary, we can state that the $F_{2}$ data for $Q^{2}\; \gtrsim \;2$
  GeV$^{2}$ are well
  described by pQCD DGLAP analyses without any sign of damping
  of the rise versus low $x$. The decrease of $F_{2}$ versus small
  $Q^{2}$ (typically $Q^{2}\; \lesssim \;1$ GeV$^{2}$) corresponds to
  the smooth transition to photoproduction which is described by
  non-perturbative models used for soft hadronic
interactions.

\section{Inclusive diffraction $(ep \rightarrow eYX)$}

\begin{figure}[ht]
 \begin{center}
 \begin{picture}(200,84)
%  had 70 before
   \put(-35,0.){\epsfig{file=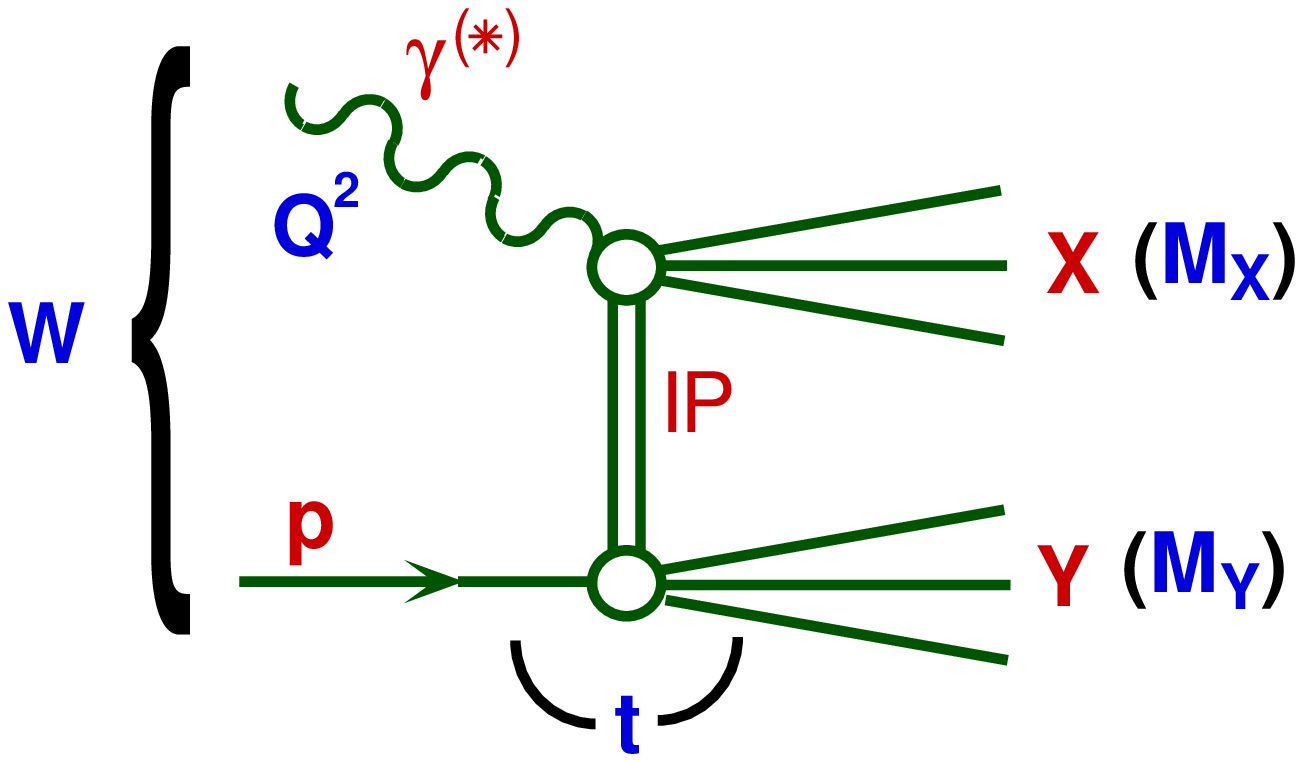,width=120pt}}
%  vorher 100
   \put(150,0.){\epsfig{file=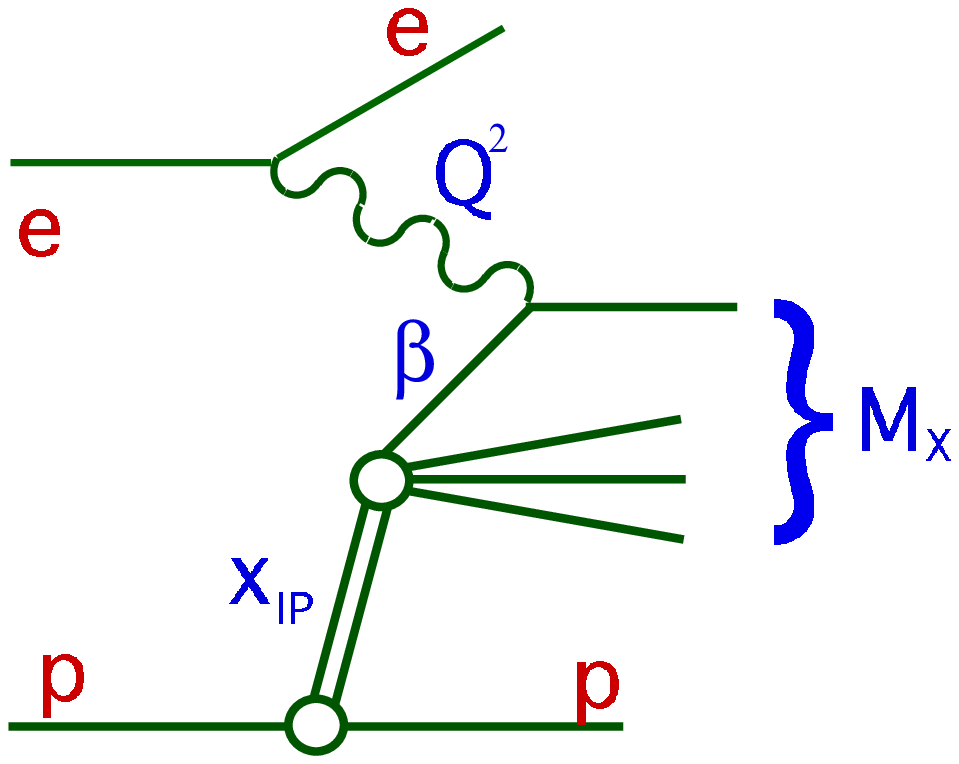,width=96pt}}
% vorher 80
   \put(-55,60.){a)}
   \put(125,60.){b)}
 \end{picture}
 \caption{Inclusive diffractive reaction  \label{XY:fig}}
 \end{center}
\end{figure}

 Interactions at HERA are called diffractive,
 when the exchanged virtual photon
$\gamma^{*}$ and the incident proton dissociate into
systems $X$ and $Y$ with colourless exchange
indicated by $\pom$ in Fig.~\ref{XY:fig}a.
 Such reactions are selected by demanding a large rapidity gap which
 then separates the systems $X$ and $Y$, or by identifying the system
$Y$
 by direct measurement of a leading proton.
 In this talk new results
 from H1~\cite{h1diff}
 are presented
%  for $Q^{2} \geq 6.5$ GeV$^{2}$ and $Y=p$.
 which are described in terms of
 a QCD analysis of the diffractive exchange.
 The data of H1~\cite{h1diff} and ZEUS~\cite{zeusdiff}
 are also compared to predictions of colour dipole models.

 The interaction can be characterised kinematically by the
 variables
\begin{equation} \label{eq:dif}
 \xpom = \frac{q . (p - p^{\prime})}{q . p} = x_{(\pom / p)}
      \;\;\;\; , \;\;\;\;
 \beta = \frac{Q^2}{q . (p - p^{\prime})} =  x_{({\rm q} / \pom)}
\end{equation}
  where $p^{\prime}$ is the 4-momentum of the leading final state proton
  (system $Y$).
 $\beta$
  can be interpreted as the momentum fraction of the exchanged
  object $\pom$ which participates in the interaction with the virtual
photon,
% (see Fig.~\ref{kine:fig},
  $x_{\pom}$ corresponds to the momentum fraction
  of the proton contributing to the colourless exchange
  (see Fig.~\ref{XY:fig}b).
 The two variables are related by  $x = x_{\pom} \beta$.

  The data are presented in terms of the diffractive structure
  function
\begin{equation} \label{eq:f2d3}
 F_2^{D(3)} (\beta, Q^2, \xpom) = \frac{\beta \ Q^4}{4 \pi \alpha^2 \ (1
- y + y^2 / 2)} \ \
\frac{{\rm d} \sigma_{ep \rightarrow eXY}}{{\rm d} \beta  \ {\rm d}
Q^2 \ {\rm d} \xpom}
\end{equation}
  defined in analogy to the fully inclusive structure function $F_{2}$
  assuming $F_{L}^{D(3)} = 0$.
   Here $y$ is the inelasticity given by
   $(p_{e}-p_{e}^{'})\cdot p/p_{e}\cdot p$.
  The results are obtained by integration over
%  the mass range $M_{Y} < 1.6$ GeV and
 $|t|=|(p-p')^{2}| < 1$ GeV$^{2}$.

  It has been shown~\cite{Trentadue:1994ka}
  that hard diffractive DIS can be factorised
  into probability distributions for partons $i$ for the case of
 an intact final
  state proton, which depend on $x_{\pom},t,x,Q^{2}$,
  and $\gamma^{*}$-parton
  cross sections $\sigma_{\gamma i}(x,Q^{2})$.
%  depending on $x$ and $Q^{2}$.
  Soft hadron phenomenology, on the other hand, suggests
  ``Regge factorisation''.
  Here a universal $pomeron$ flux $f_{\pom/p}(x_{\pom},t)$ is assumed.
%  depending on $x_{\pom}, t$ only.
  The cross section can then be written as
\begin{equation} \label{eq:res}
\sigma (\gamma^* p \rightarrow X p)
\sim
f_{\pom / p}(\xpom, t) \otimes
\sum_i f_{i / \pom} (\beta, Q^2)
\otimes
\hat{\sigma}_{\gamma^* i} (\beta, Q^2)
\end{equation}
  where the $f_{i}$, in the spirit of Ingelman and
Schlein~\cite{Ingelman:1985ns},
  are the densities of the partons which interact with the virtual
photon.

\begin{figure}[ht]
 \begin{center}
 \begin{picture}(200,268)
%   \put(0,0.){\epsfig{file=figs/fig7b.eps,width=0.55\textwidth}}
   \put(-50,0.){\epsfig{file=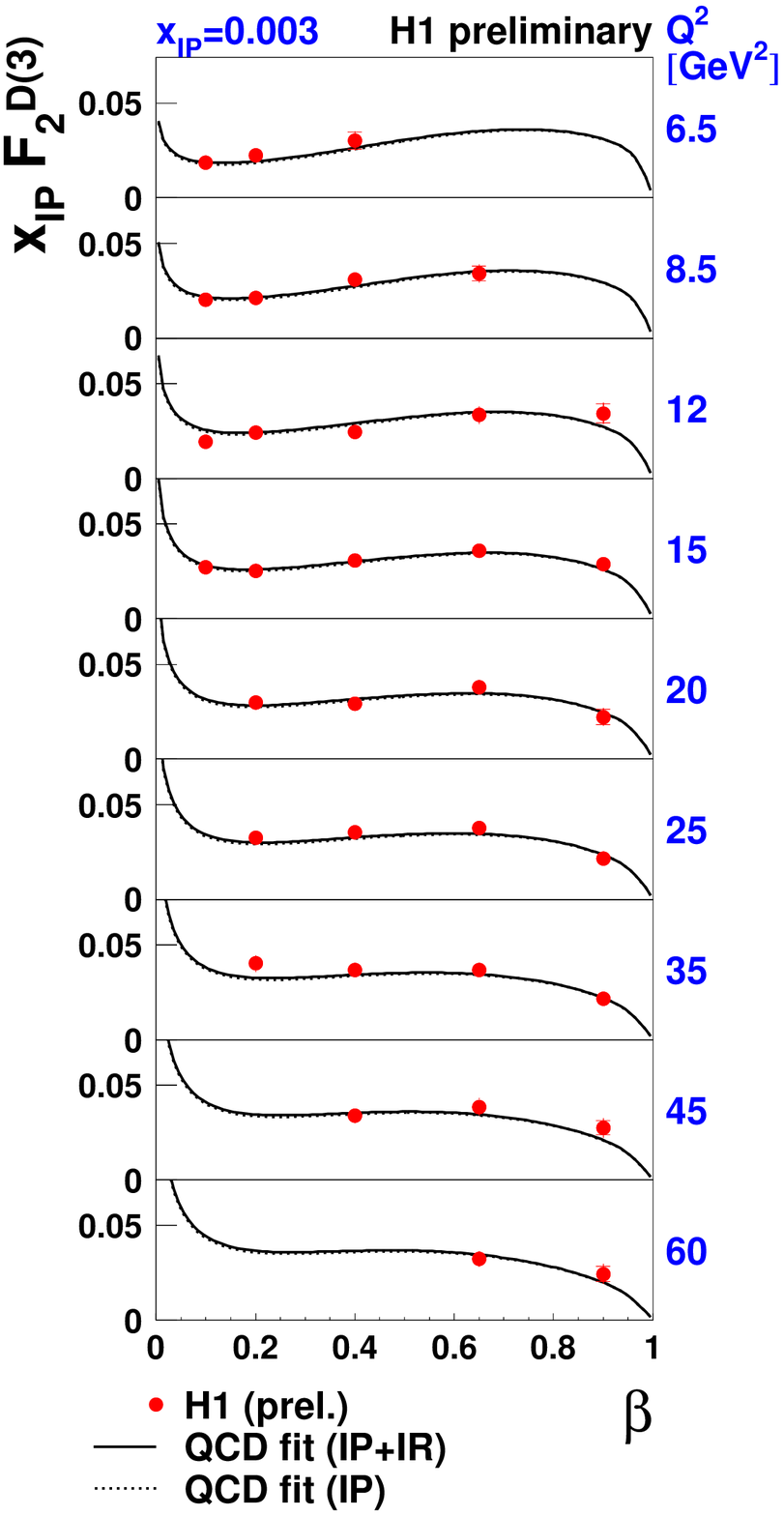,width=140pt}}
   \put(110,0.){\epsfig{file=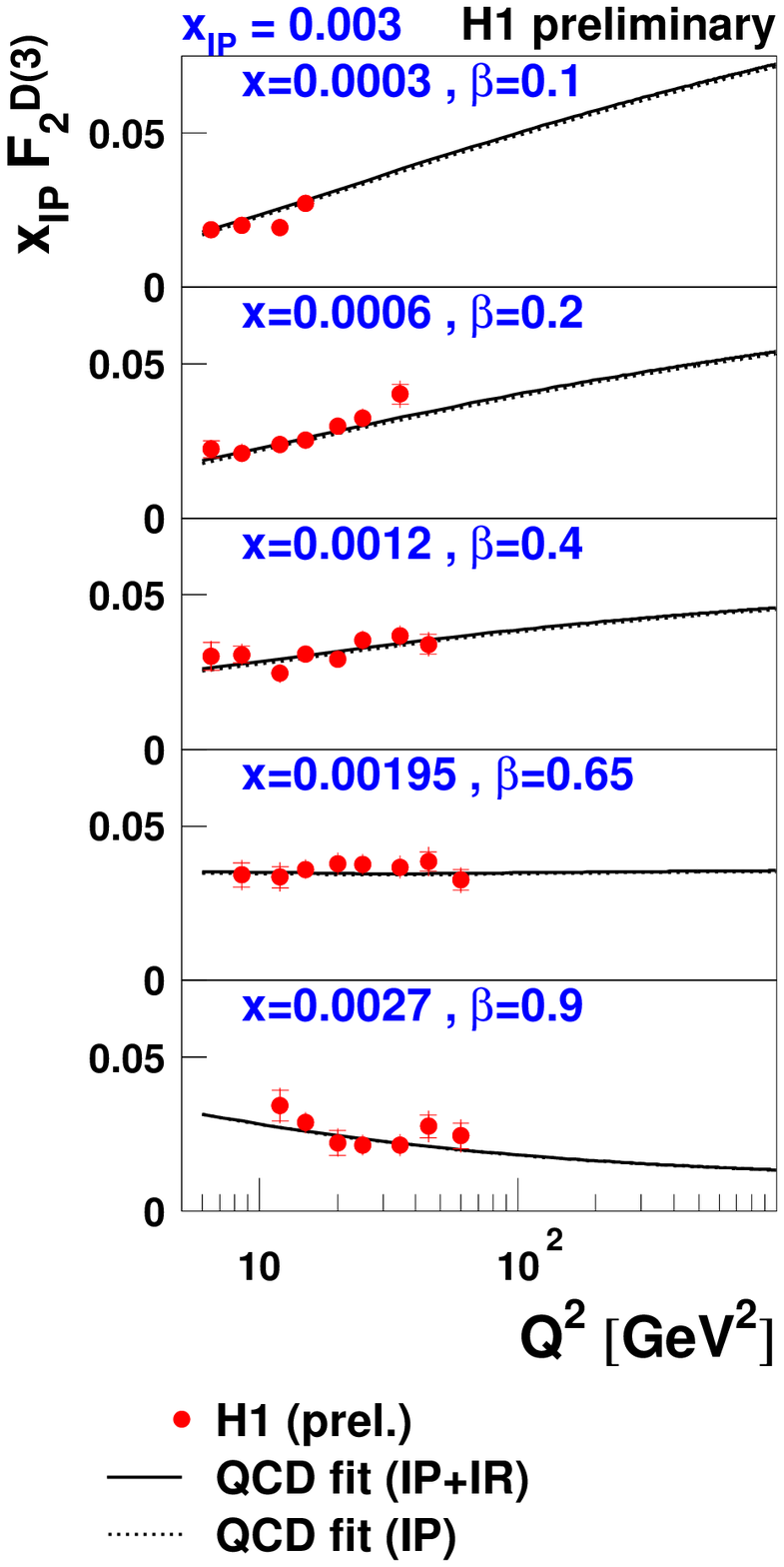,width=140pt}}
 \end{picture}
   \caption{ $\xpom F_{2}^{D(3)}$ as function of $\beta, Q^{2}$
    and resolved $\pom$ fit ($6.8\%$ normalisation error of data not shown).
   \label{betaqcd:fig}}
 \end{center}
\end{figure}

  The precise H1 data (Fig.~\ref{betaqcd:fig})~\cite{h1diff},
 which cover the range
  $6.5 \leq Q^{2} \leq 120$ GeV$^{2}$ and $0.04 \leq \beta \leq 0.9$,
  are well described by a fit assuming Regge factorisation, with the
   result
 $\alphapom(0) = 1.173 \pm 0.018 \mathrm{(stat.)}
                     \pm 0.017 \mathrm{(syst.)}
                     ^{+ 0.063}_{-0.035} \mathrm{(model)}$.
   Compatible results are obtained if the fit is
   performed in two separate $Q^{2}$ ranges.
%   These results are shown
%   in Fig.\ref{alpha:fig} together with other results from H1 and ZEUS.

%\begin{figure}[ht]
%\begin{center}
%%\figurebox{20pc}{15pc}{} % to have a box alone
%\epsfbox{figs/alphanew.eps} % postscript image file name
%\caption{alpha  \label{alpha:fig}}
%\end {center}
%\end{figure}

%\begin{figure}
%\begin{center}
%\epsfig{file=figs/alphanew.eps,bbllx=0pt,bblly=16pt,bburx=262pt,bbury=324pt,%

%angle=0,width=11cm,clip=}
%\end{center}
%\end{figure}
   We can state (see also~\cite{zbpc}):   
\begin{itemize}
\item
 at $Q^{2} \gtrsim 1$ GeV$^{2}$, $\alpha_{\pom}(0)$ is above the soft
   pomeron expectation~\cite{Donnachie:1992ny} of 1.08, 
\item
 $\alpha_{\pom}(0)$ seems to grow slower in diffraction than
    in inclusive DIS
  (see eq.\ref{eq:f2p}).  
\end{itemize}

In spite of
 the latter point, the energy dependencies are similar
for diffractive and inclusive DIS. This corresponds to the different
expected dependencies
$\sim~x^{2-2\alphapom(t)}$ and
$\sim x^{1-\alphapom(0)}=x^{-\lambda}$
% respectively (see sect's 1 and 2).
 respectively (see sections 1 and 2).
%cosmetic

 The results of a pQCD fit in terms of pomeron pdfs  
 to the $\beta, Q^{2}$ dependence of the data 
% A DGLAP fit to the $\beta, Q^{2}$ dependence of the data in terms
 are shown in
 Fig.~\ref{betaqcd:fig} for the case $x_{\pom} = 0.003$.
%\begin{figure}[ht] \unitlength 1mm
 The $\beta$ dependence is rather flat,
 the scaling violations in $\ln Q^{2}$ are positive
 up to large
        $\beta$, with the consequence that a large gluon component
  is required for the pdfs of the pomeron up to large $\beta$.

 Quite differently from the approach of resolved pomeron models
 discussed above,
 colour dipole models
 (e.g.~\cite{Buchmuller:1999jv,Golec-Biernat:1999js,Golec-Biernat:2001mm})
 aim to describe
 the long distance features of the interaction by modeling
 an effective cross section for the interaction of $q \bar q$ pairs or
 $q \bar q g$ systems with the proton. Seen in the proton rest frame,
 these colour dipoles are produced perturbatively at large
 longitudinal distances
  $\sim 1/m_{p} x$, well outside the proton for small $x$.
  The effective transverse size of these fluctuations is
  $<r^{2}> \sim 4/Q^{2}$.
  In the model of Golec-Biernat and
  W\"usthoff~\cite{Golec-Biernat:1999js} saturation
  of the dipole cross section at large radii,
  i.e. small $Q^{2},
  p_{t}^{2}$, is introduced which at small $x$ is reached already
  at smaller radii.
%  With such models
%  non-perturbative phenomena are investigated by the tentative
%introduction of
%  dipole-$p$ cross sections which lead to realistic results.

\begin{figure}[ht]
 \begin{center}
 \begin{picture}(200,268)
   \put(-50,0.){\epsfig{file=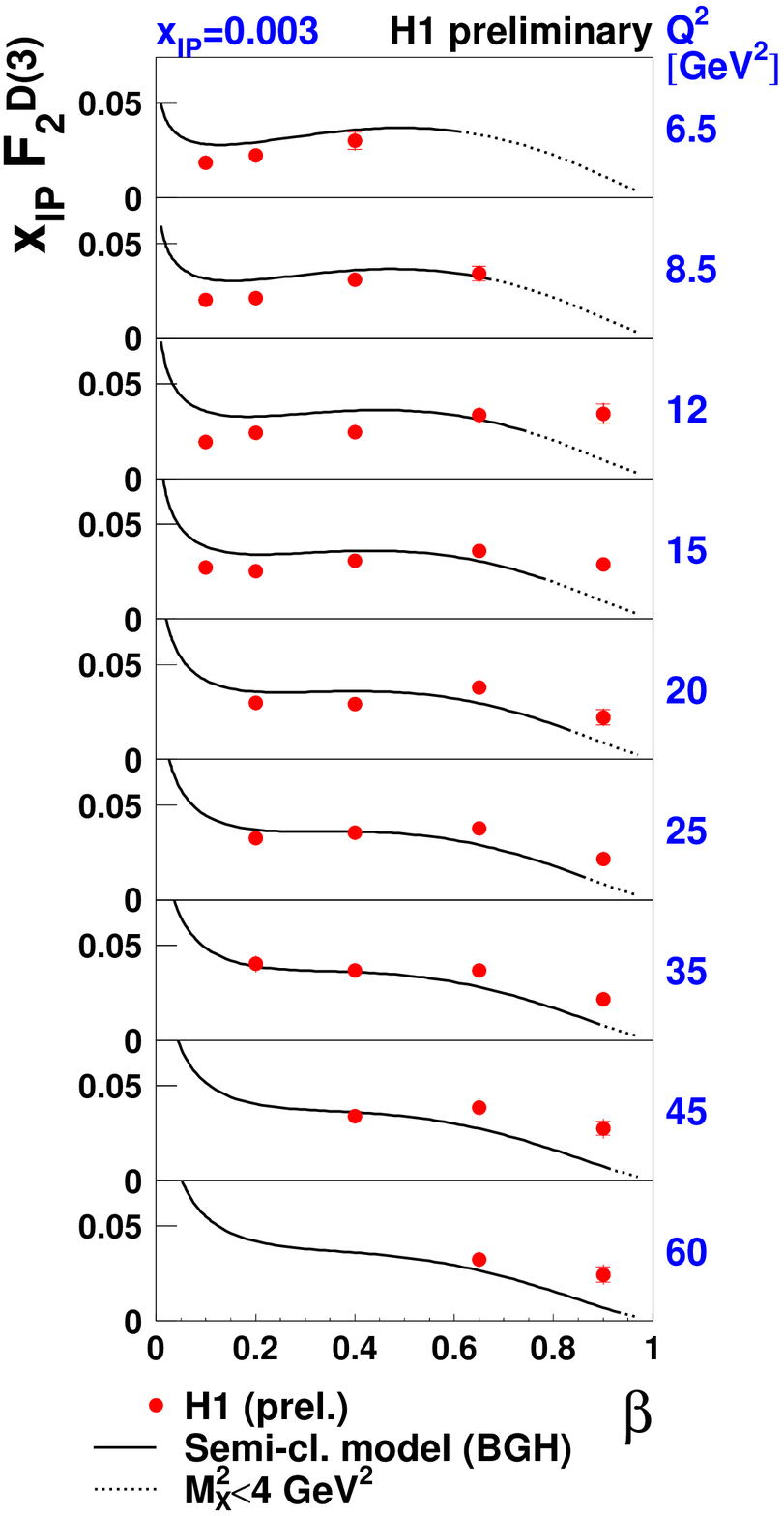,width=140pt}}
   \put(110,0.){\epsfig{file=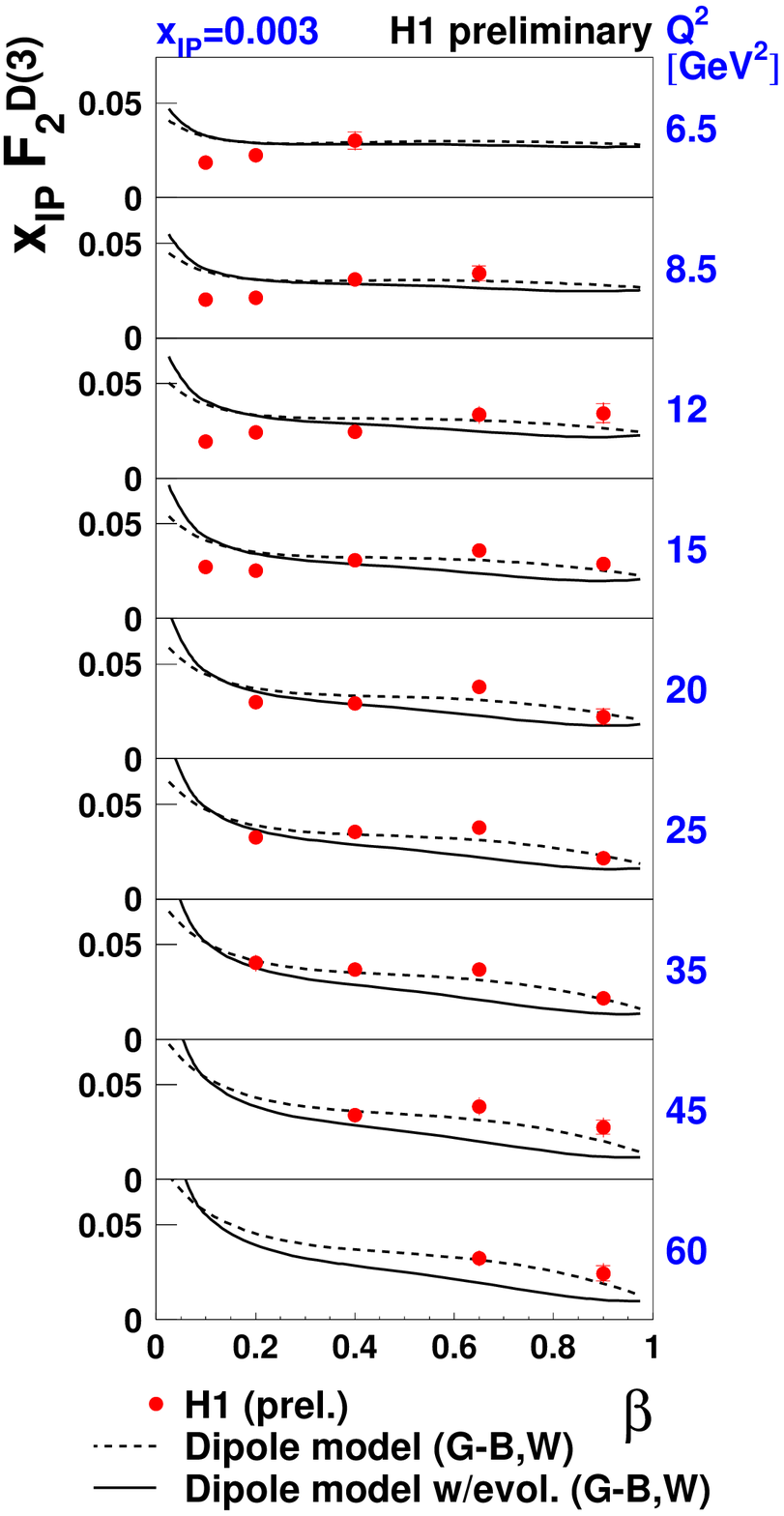,width=140pt}}
 \end{picture}
   \caption{$\xpom F_{2}^{D(3)}$ as function of $\beta, Q^{2}$ and
   colour dipole models [18,19,20].
% hand ref
   \label{betacolour:fig}}
 \end{center}
\end{figure}

  Two colour dipole models~\cite{Buchmuller:1999jv,Golec-Biernat:1999js}
  are compared with the measured $\beta$ distributions
  of $x_{\pom} F_{2}^{D(3)}$ for different $Q^{2}$ in
  Fig.~\ref{betacolour:fig}.
 The general features of the data  are
 described by the models, but both are above the data at low $\beta$,
 low $Q^{2}$. The saturation model~\cite{Golec-Biernat:1999js}
 contains a higher twist contribution
 at large $\beta$, but no improvement is achieved by
 a more advanced version
 involving $Q^{2}$ evolution~\cite{Golec-Biernat:2001mm}.

\begin{figure}[ht]
 \begin{center}
 \begin{picture}(200,312)
   \put(-65,0.){\epsfig{file=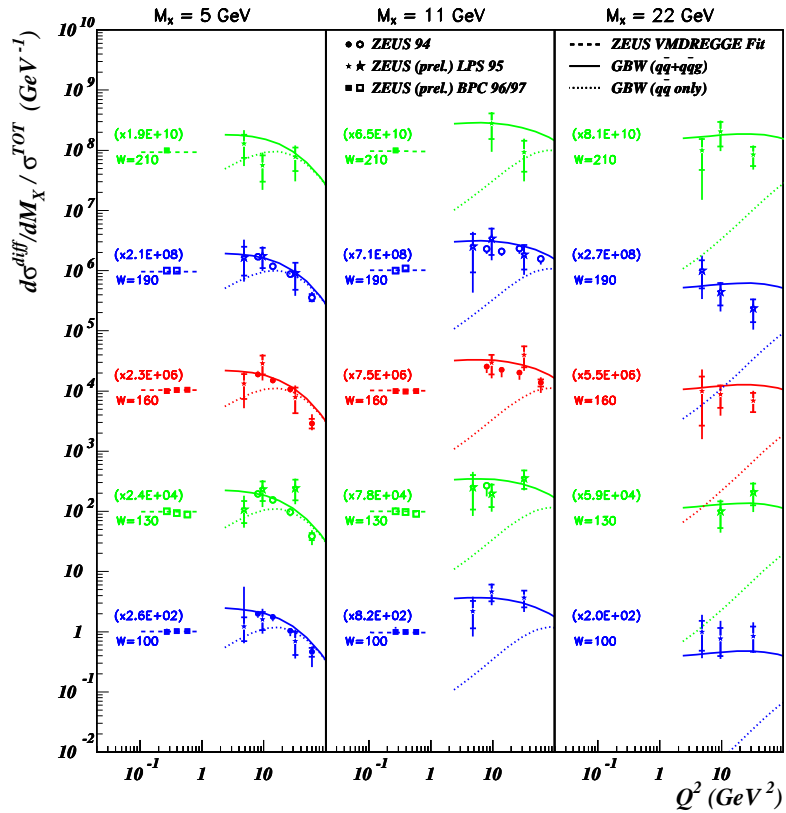,width=300pt}}
 \end{picture}
\caption{Ratio of diffractive and inclusive cross sections in 3 $M_{X}$
 ranges as function of $Q^{2}$ compared with the colour
% dipole model~\cite{gw}. \label{zeusdiff:fig}}
 dipole model [19].  \label{zeusdiff:fig}}
%  hand ref
 \end{center}
\end{figure}

 The colour dipole models successfully predict the similarity of the
  energy dependencies of diffractive and fully inclusive $ep$ DIS
  as shown in Fig.~\ref{zeusdiff:fig} where the ratio of the
  corresponding cross sections~\cite{zeusdiff} as
  function of $W$ is
  compared with the saturation model~\cite{Golec-Biernat:1999js}.
 The $q\bar q g$ component in the model is dominant at large $M_X$.

% Final state properties of inclusive diffractive
% DIS~\cite{h1zfinal} are quite well
% described by Monte Carlo models based on a resolved $\pom$~\cite{jung}
%or photon
% dissociation into colour dipoles~\cite{ridisat}.

 The main results on inclusive diffraction can be summarised as
 follows:          
\begin{itemize}
\item
 The data are consistent with Regge factorisation. 
\item
 The effective $\alpha_{\pom}(0)$ is at large $Q^{2}$ larger than
        expected from soft hadronic
  interactions.   
\item
 The scaling violations $dF_{2}^{D3}/dQ^{2}$ are positive up to large
      $\beta ( > 0.4)$ which
 is interpreted in DGLAP QCD analyses as a
        large gluon content in pomeron
 pdfs. 
\item
  Colour dipole models give a reasonable description of the main
  features of
 the data.
\end{itemize}
\section{Exclusive Diffractive Reactions $(ep \rightarrow eVp,
     ep \rightarrow e\gamma p)$}
   Exclusive  vector meson and $\gamma$ production
  is particularly useful
   to gain understanding of diffractive processes.
   In these reactions the
   transverse interaction radius can be varied systematically by
   choosing different  $Q^{2}$ and particle mass $M_{V}$.

\begin{figure}[ht]
 \begin{center}
 \begin{picture}(200,178)
% vorher 148
   \put(-105,15.){\epsfig{file=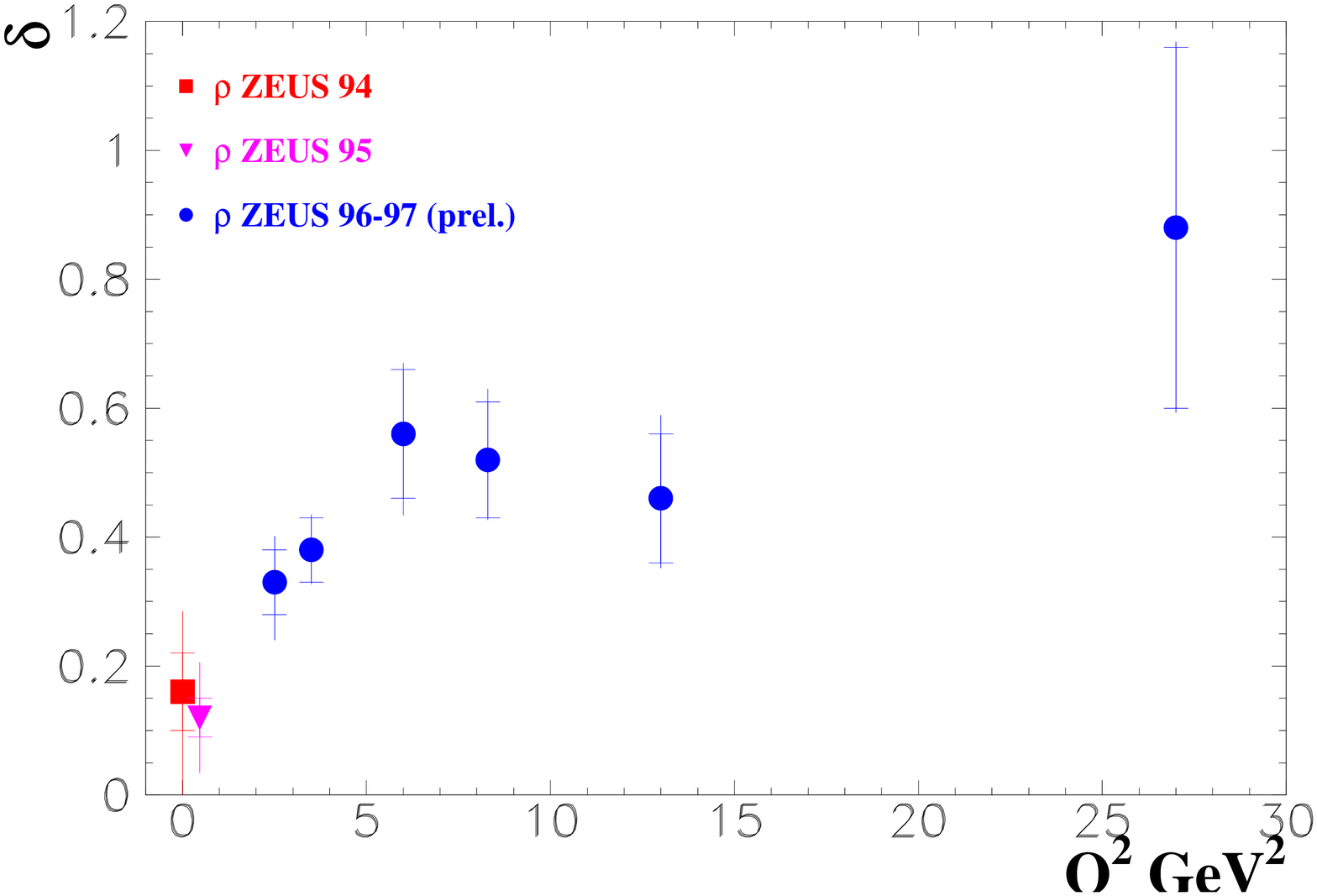,width=204pt}}
%   vorher 170
   \put(130,0.){\epsfig{file=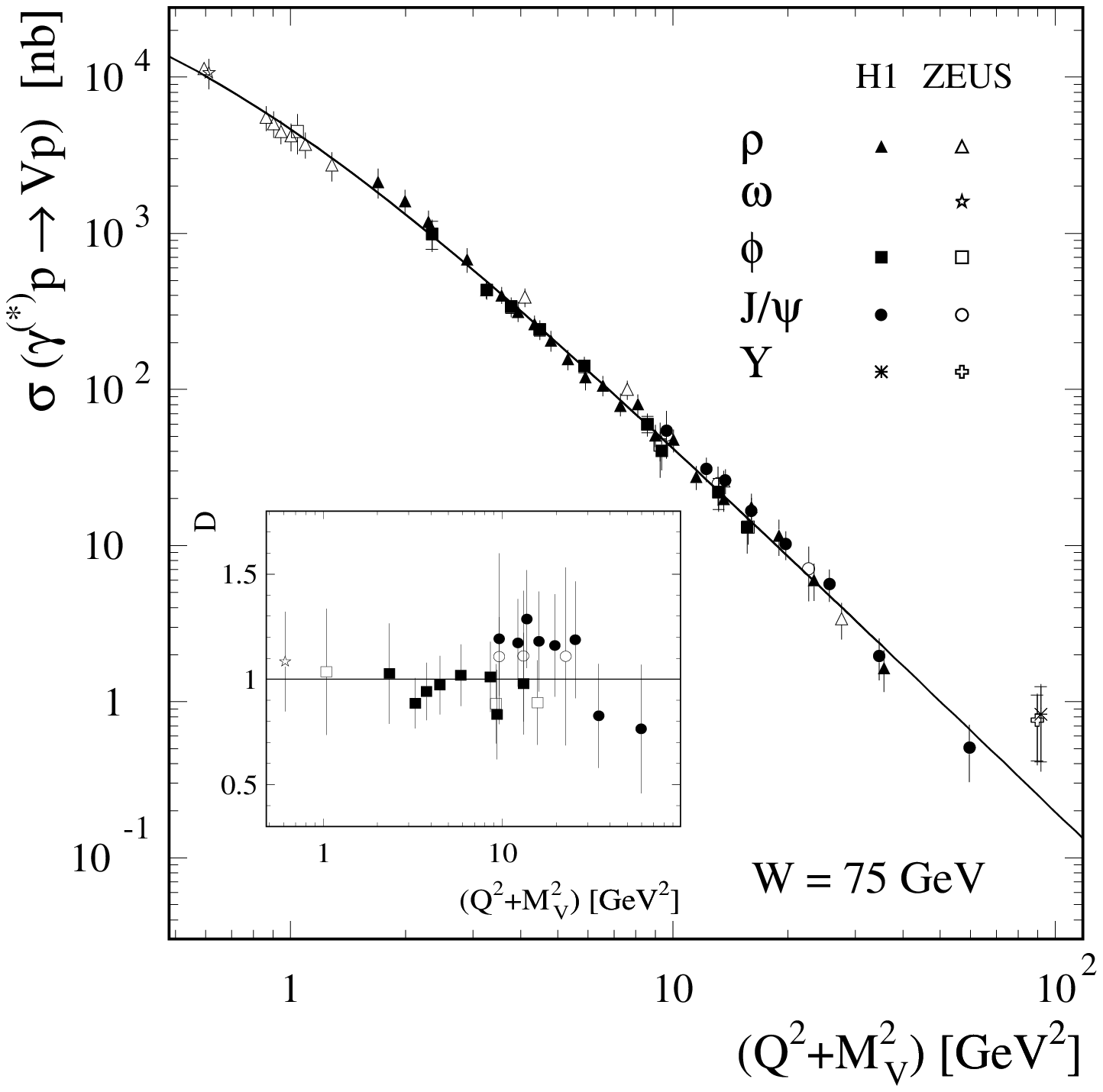,width=192pt}}
%  vorher 160
   \put(-11,137.){a)}
   \put(223,137.){b)}
 \end{picture}
   \caption{ a) Exponent $\delta$ of from a fit $\sim W^{\delta}$
    to $\rho$ production vs. $Q^2$. b) Vector meson production
    as function of $Q^{2} + M_{V}^{2}$
    scaled by SU(4) factors $\rho^0 :\omega:\Phi:J/\Psi = 9:1:2:8$.
   \label{vmscale:fig}}
 \end{center}
\end{figure}

   The energy ($W$)
   dependence gets steeper with both, increasing particle mass
   in photo production (see e.g.~\cite{Adamczyk:2001fh})
%   (Fig.~\ref{vmphoto:fig})
 and also with increasing $Q^{2}$ as
   demonstrated for $\rho$ production in Fig.~\ref{vmscale:fig}a
   which shows the exponent $\delta$
   for fits to ZEUS data~\cite{zrhoq2} according
   $\sigma \sim W^{\delta}$. In both cases the rise is steeper
   when the reactions get less peripheral.
%\begin{figure}[ht]
%\begin{center}
%\epsfxsize=20pc % will enlarge or reduce the postscript figures
%\epsfbox{figs/rhowdelta.eps} % postscript image file name
%\caption{rho $\delta , W^\delta$  \label{zrhoq2:fig}}
%\end {center}
%\end{figure}
   In fact an approximate scaling behaviour of vector meson production
   with the effective mass squared $Q^{2} + M_{V}^{2}$ can be observed
   (Fig.~\ref{vmscale:fig}b). It should be
   noted that the figure shows the
   sum of longitudinal ($\sigma_{L}$) and transverse ($\sigma_{T}$)
   cross sections which are known to behave differently, 
%   More recent data~\cite{zrhopsi} in the range
%   $9.6 \leq Q^{2} + M_{V}^{2} \leq 25.6$ GeV$^{2}$
%   $10\;\lsim \; Q^{2} + M_{V}^{2}\; \lsim \; 25$ GeV$^{2}$
%   indicate that 
%   $J/\psi$ production is $\approx 30\%$ above the expectation
  and further, $J/\psi$ production seems to be somewhat above the
  expectation from the scaling behaviour, 
 as seen in a more recent analysis~\cite{zrhopsi}. 
%   from $\rho$ production assuming the scaling behaviour, as seen 

   The $t$ dependence $\sim exp(-b|t|)$ of diffractive
   vector meson production yields a direct measure of the transverse
   size $r$ of the interaction region ($b$ $\sim <r^2>$).
   The HERA data (e.g.~\cite{mellado})
%   (Fig.~\ref{bslope})
   show indeed that $b$ decreases for
   increasing $Q^{2} + M_{V}^{2}$.
   It is however an interesting question,
   whether in vector meson production at large
   $Q^{2}$  or $M_{V}^{2}$,
   the effective size of the interaction region increases with energy
   as it is observed in soft hadronic interactions (``shrinkage'' of the
      $t$-slope $b$).
   The available data on the energy dependence of $\rho$
 and $J/\psi$
production at
   different $t$ indicate a slight increase of the interaction
   region with increasing energy~\cite{Adloff:2000vm}, but less than expected
from
   soft hadronic interactions.

\begin{figure}[ht]
 \begin{center}
 \begin{picture}(200,149)
%    vorher 142
   \put(-60,-14.){\epsfig{file=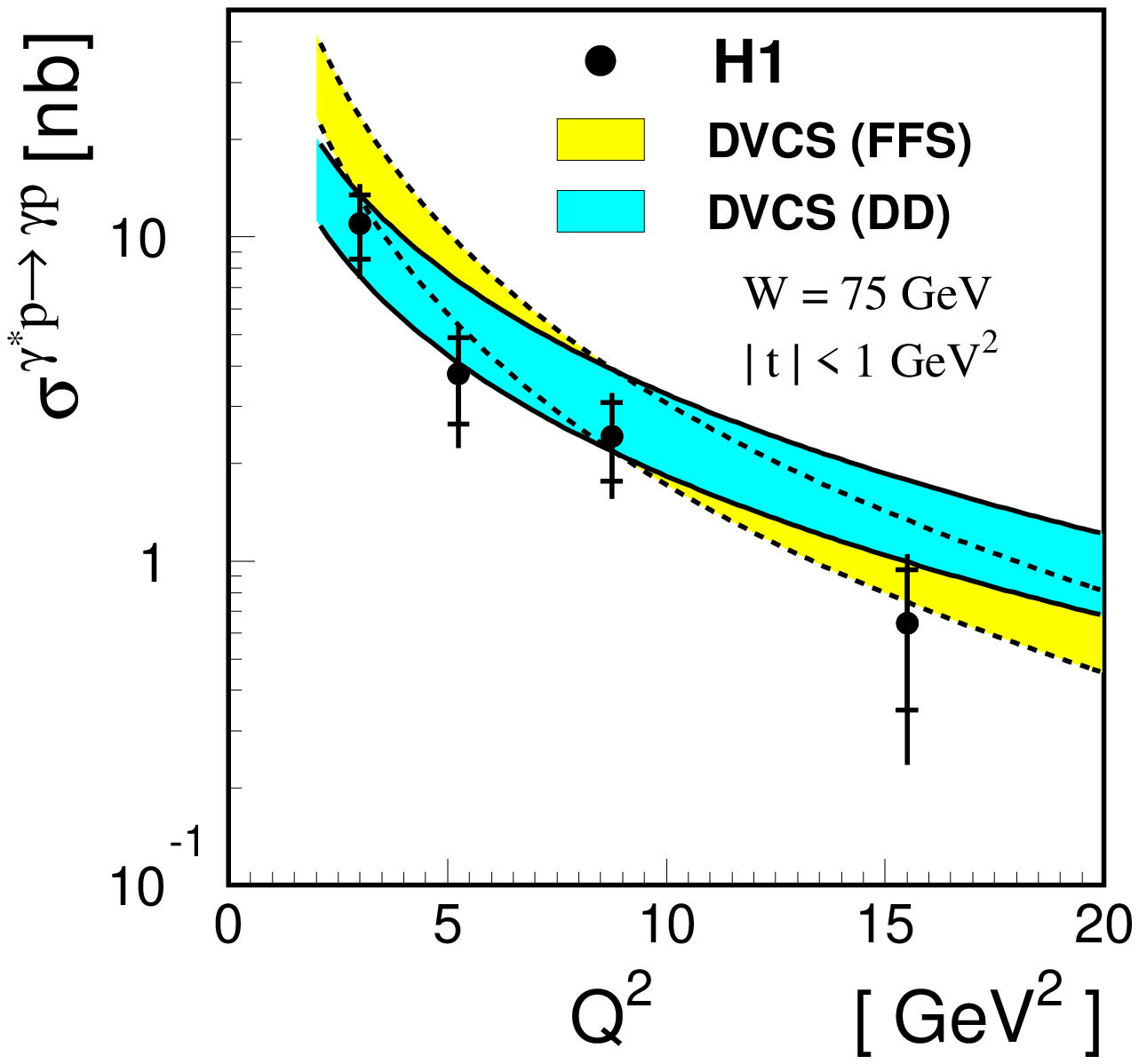,width=157pt}}
   \put(130,1.){\epsfig{file=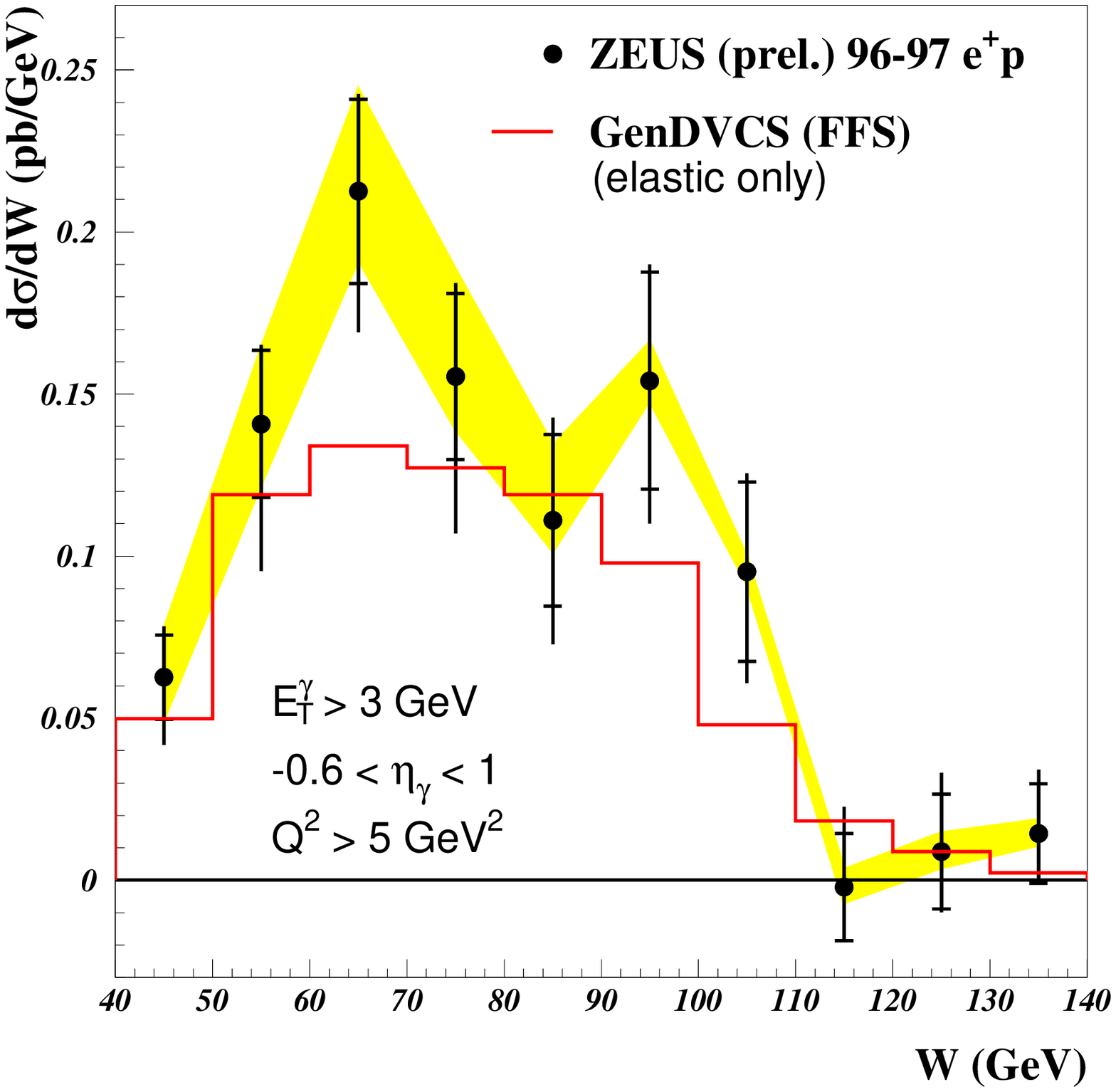,width=147pt}}
 \end{picture}
   \caption{DVCS results as function of $Q^{2}$ and $W$
%   compared with a leading order QCD model (FFS~\cite{ffs}) and
   compared with a leading order QCD model (FFS [30]) and
%  $\pom$ and Reggion exchange (DD~\cite{Donnachie:2001px}). \label{dvcs:fig}}
   $\pom$ and Reggion exchange (DD [31]). \label{dvcs:fig}}
%   two hand references
 \end{center}
\end{figure}

  First data on deep inelastic Compton scattering
  $\gamma ^*p \rightarrow \gamma p$ (DVCS) have been
   presented
 recently~\cite{Airapetian:2001yk,Adloff:2001cn,Stepanyan:2001sm,dvcsz}.
   DVCS is a particularly clean process of light-quark scattering.
   The proton vertex is similar as in vector meson production,
   given in the simplest case by two  exchanged gluons or quarks,
   but free of the complication by quark wave functions of the vector
meson
   states. The simplicity of the process may give
   better access to the
   ``skewed'' or generalised parton density distributions in the proton
   than vector meson production.
   H1~\cite{Adloff:2001cn} and ZEUS~\cite{dvcsz}
   subtracted the contribution of the purely electromagnetic
   Bethe Heitler (BH) reaction, exploiting its different angular
distribution.
   The resulting $Q^2$ and $W$ distributions are
   reasonably described by a leading order QCD model
   including non-perturbative elements~\cite{Frankfurt:1998at}
  and a model based on
   pomeron and Reggeon
   exchange~\cite{Donnachie:2001px} (Fig.\ref{dvcs:fig}).
   The HERMES and CLAS collaborations determined the interference with
the
   BH process by measuring the beam spin asymmetry.
   This field is rapidly developing:
   Next to leading order QCD calculations are meanwhile
   available~\cite{Freund:2001hd}
   and  more data from HERA are to be expected, in particular on the
   DVCS-BH interference.
\section{Conclusions}
\begin{itemize}
\item
% $\bullet$
 HERA provides information on the transition from short
  to long distance
%   \hspace*{7pt}
   phenomena. 
\item
 The rise of $F_2$ versus small $x$ at $Q^2 \gtrsim 2$ GeV$^2$ shows
no effect of saturation
            at present energies.   
\item
  The colour singlet exchange in diffractive reactions can be described
  by $\pom$
 parton densities (``resolved $\pom$''). The data require
  a large gluon component.     
\item
 The energy dependence of vector meson production is steep when 
 objects
 of small
 transverse size ($J/\Psi,\; \rho^0$ at high $Q^2$) are involved,
 but comparable
      to that of soft hadronic interactions in other cases. 
\item
 First results on deeply virtual Compton scattering (DVCS) are consistent
 with calculations based on leading order QCD and non-perturbative
 model
 assumptions.
 Future data will supply more information on
 generalised
 parton distributions.
\end{itemize}
\section*{Acknowledgments}
I am grateful to P. Schleper and M. Derrick for comments on the manuscript.

\end{document}

%% file: comdef.tex
\newcommand{\pom}{I\!\!P}

\newcommand{\xpom}{x_{\pom}}

\newcommand{\alphapom}{\alpha_{_{I\!\!P}}}